\documentclass[12pt]{JHEP3}

\usepackage{epsfig,amssymb,cite,multirow} 
\usepackage{amsmath}

\usepackage{amscd}
\usepackage{amsmath,graphicx}
\usepackage{xypic}
\usepackage[matrix,arrow,curve]{xy}
\usepackage{graphicx}
\usepackage{verbatim}
\usepackage{epsf}
\usepackage{latexsym}
\usepackage{amsmath,amsfonts,amssymb,amsthm}
\usepackage{amsmath,amsthm}

\def\bseq{\begin{subequation}}  
\def\eseq{\end{subequation}}
\def\bsea{\begin{subeqnarray}}  
\def\esea{\end{subeqnarray}}

\newcommand{\bbox}{\lower.2ex\hbox{$\Box$}}

\newcommand{\beq}{\begin{equation}}
\newcommand{\eeq}{\end{equation}}
\newcommand{\bea}{\begin{eqnarray}}
\newcommand{\eea}{\end{eqnarray}}
\newcommand{\ena}{\end{eqnarray}}

\newcommand {\non}{\nonumber}

\newcommand{\Tr}{{\rm Tr}}

\newcommand{\be}{\begin{equation}}
\newcommand{\ee}{\end{equation}}

\title{\begin{center} 
 On the exact $R$ charge for $\mathcal{N}=2$ CS theories
\end{center}}
\author{
Antonio Amariti$^{1,a}$,
\\~
\\
$^1$Department of Physics, University of California\\
San Diego La Jolla, CA 92093-0354, USA
\\~\\
$^a$\email{antonio.amariti@physics.ucsd.edu} \\
}

\abstract{Recently it was argued that the exact $R$ charge for three dimensional
$\mathcal{N}=2$ supersymmetric field theories extremizes the 
partition function localized on $S^3$. In this paper we check this conjecture 
by computing the $R$ charge for $SU(N)_k$ YM CS
gauge theories at large $k$ for many representations,
and we test the agreement with the perturbative  results.
}

\preprint{UCSD PTH-11-03}

\begin{document}

\section{Introduction} \label{sec0}

In the IR fixed point of a superconformal field theory the $U(1)_R$ charge 
usually differs from its UV value. Indeed the UV $R_0$ charge mixes with the global non-R abelian 
flavor symmetries $F_i$ during the RG flow. At the fixed point the exact R charge
that belongs to the superconformal algebra is 
a combination of these charges. 
In some cases, like $\mathcal{N}=1$ $4d$ SQCD in the conformal window, the constraint arising from the superpotential
and the requirement of anomaly cancellations  determine the exact $R$ charge at the IR fixed point. 
Anyway in many other cases these constraints are not enough and further informations are necessary.
The problem of finding the exact IR R symmetry among all
the trial combinations  $R_t=R_0 +  \alpha_i F_i$ has been solved for $\mathcal N = 1$ 
$d=4$ superconformal field theories in  \cite{Intriligator:2003jj}.
It was shown that the combination of $\alpha_i$ that gives the
exact $R$ charge at the fixed point maximizes the
conformal anomaly. This is associated with the Euler term in the trace anomaly of the 
stress tensor, $\langle T_{\mu}^{\mu}\rangle \simeq c ($Weyl$)^2 - a ($Euler$)$,
where $a \simeq 3$Tr$R^3-$Tr$R$.
This result was shown by using the condition that the global $U(1)_R $ is free of ABJ anomalies
(or equivalently the condition that the exact NSVZ  beta function \cite{Novikov:1983uc} vanishes for all the gauge groups)
and that the $U(1)_R$ 't Hooft anomalies determine the central charges $a$ and $c$ \cite{Anselmi:1997ys}. By using the 't Hooft anomaly matching it was indeed possible to compute these central charges in terms of the weakly coupled UV theory, and not from the
 IR physics, usually  strongly coupled.

The same problem of finding the exact $R$ charge at the superconformal fixed point
can be formulated in $\mathcal{N}=2$ $d=3$. For higher amount of supersymmetry 
the $U(1)_R$ symmetry does not mix with the other 
abelian symmetries of the theory, and $R_{IR}=R_{UV}$.
 Unfortunately in the $\mathcal{N}=2$ case the $a$-maximization
cannot be performed, because the trace anomaly of the stress tensor vanishes in 
three dimensions. 
In \cite{Barnes:2005bm} it was argued that in $d$ dimensional superconformal gauge theories the exact $R$ charge minimizes the coefficient $\tau$ of the $R$-current two point function
\be
\langle J_{R}^{\mu}(x) J_{R}^{\nu}(y) \rangle \simeq \tau_{RR} \left(\partial^2 \delta^{\mu \nu} - \partial^\mu \partial^\nu \right) \frac{1}{(x-y)^{2(d-2)} }.
\ee
Anyway in three dimensional theories this quantity receives quantum corrections and 
a perturbative analysis is necessary.

Recently a different approach to this problem has been proposed in \cite{Jafferis:2010un}.
It was argued that in three dimensions the role of the central charge $a$ is 
played by the three sphere partition function $\mathcal{Z}$.
This procedure of putting the theory on a sphere, which is called localization,
was already applied on $d=4$ $\mathcal{N}=2$ supersymmetric gauge theories in \cite{Pestun:2007rz}.
It consists of adding some $Q$-exact operator (where Q represent the supersymmetry generator on
the sphere), such that the path integral does not change but the one loop determinants are exact.
This is the key point of the computation. Indeed even if the $R$ charge mixes with the abelian flavor symmetries, the partition function 
at one loop contains all of the information of this mixing. 
Recently in $d=3$ the localized partition functions on  $S^3$
have been investigated for $\mathcal{N}\geq 3$ supersymmetry \cite{Drukker:2010nc,Herzog:2010hf},
 where it was show that the $N^{3/2}$ entropy scaling expected from the gravitational side of the 
 Ads$_4/$CFT$_3$ correspondence \cite{ABJM} can be computed in field theory.
 Moreover the study of the $\mathcal{N}=2$ case was started in \cite{Kapustin:2009kz},
 in theories where there are not quantum corrections.  The novelty of \cite{Jafferis:2010un}
 is that the quantum corrected partition function computed on  $S^3$  is extremized
 by the exact IR $R$ symmetry. 
In \cite{Martelli:2011qj,Cheon:2011vi,Jafferis2}  this conjecture was strengthened by computing the partition function of some
 $\mathcal{N}=2$ CS toric quiver gauge theories. Indeed in that case it is possible to compute the exact 
 field theory $R$-charge by minimizing the volume of the Sasaki Einstein manifold \cite{zmin}.
Actually in four dimensions this function is matched  with the conformal anomaly $a$ even before the 
 extremization \cite{Butti:2005vn}.  The authors showed that a similar relations between the Sasaki Einstein 
 volume and the partition function holds in three dimensions at large $N$.

In this paper we study the conjectured $\mathcal{Z}$ extremization in CS matter theories at large CS level $k$.
 In \cite{Jafferis:2010un} 
 the $U(1)_k$ and $SU(2)_k$ CS gauge theories with flavor were discussed, and the
agreement with the field theory perturbative analysis was checked.
The localization procedure reduces the path integral to a matrix integral over the Cartan subalgebra
of the Lie algebra the $U(1)_k$ and $SU(2)_k$  and the integrals 
are over a single variable. 
Here we check the conjecture by computing the partition function of non abelian CS theories
with higher dimensional Cartan subalgebra, with matter fields 
in the $R+\overline R$ representation.
We study   $SU(N)_k$ 
theories and we see that the computation agree with the perturbative results discussed in field theory.
For large $k$ the matrix integral are dominated by a gaussian factor $e^{k \Tr u^2}$, where $u_i$
are the eigenvalues in the Cartan basis. 
The computations is  performed by expanding the partition function at the saddle point,
$u\sim0$.
This expansion simplifies the integral by giving an overall Vandermonde determinant coming
from the one loop determinant of the vector multiplet.
By writing this Vandermonde determinant in terms of orthogonal polynomials we have been able to 
compute the exact partition function at the lowest order in the  't Hooft coupling.
\\
\\
In section \ref{sec1} we review the basic aspect of the partition function. In section \ref{sec2}  we review the field theory 
results.  In section \ref{sec3} explain the strategy of the computation of the partition function for the $SU(N)_k$ theories,
and we check that the $R$ charge evaluated with this method agrees with the two loop results for fundamental, adjoint, 
symmetric and antisymmetric representations.
Then we conclude. In appendix \ref{sec6} we review the conventions we used on Lie algebra, while
in appendix \ref{sec7}  we summarize some results for the integration of the orthogonal polynomials.

\section{The partition function  localized on $S^3$} \label{sec1}

As explained in the introduction the exact $R$ charge is computed by extremizing  the partition function
of the localized $\mathcal{N}=2$ action on $S^3$ of a 
YM CS theory with matter fields in the $R$ representation of the gauge group.
The computation of the partition function on $S^3$ follows from the localization techniques of \cite{
Nekrasov:2002qd}.
This technique was first used by \cite{Pestun:2007rz} for the computation of the partition function of $S^4$ for $\mathcal{N}=2$ 
supersymmetric gauge theories in four dimensions. In three dimensions the first computation was performed 
in $\mathcal{N}\geq 3$ by \cite{Kapustin:2009kz}.
Then in \cite{Jafferis:2010un,Hama:2010av} this computation was extended to the $\mathcal{N}=2$ case, in which the partition function
explicitly depends on the $R$ charge, which is not trivial, because the $R$-symmetry is abelian, and some regularization
is necessary.
In the case of a YM-CS in $\mathcal{N}=2$ three dimensional models they found that the relevant contribution to the partition function is 
\be \label{partf}
\mathcal{Z}   \simeq \int \prod_{i=1}^{N} du_i e^{ \pi i k \Tr_{F} u^2} {\det}_{\text{adj}}\left( \sinh(\pi u)\right) {\det}_R e^{l(1-\Delta+i u)}
\ee
The different terms in (\ref{partf}) are 
\begin{itemize}
\item 
the rk$_G$ integration variables over the
gauge group $G$. They are written in terms of the eigenvalues $u_i$ which 
correspond to the weight of the fundamental representation.
These variables correspond to the scalars $\sigma$ in the vector multiplet.
They transform in the adjoint representation of the gauge group, and this allows
the intagrals to be taken over $\mathbb{R}$.

\item the determinant over the adjoint is explicitly 
$\prod_{i<j} \sinh^2 (\pi \rho_{ij}(u))$
and it is related to the  one loop contribution of every vector multiplet
\footnote{We are omitting a factor $2$, which is overall and does not affect the
computation of the $R$ charge.}.
The weights $\rho_{ij}(u)$ correspond to the roots of the algebra because the vector multiplet
is in the adjoint representation;
\item
the term $e^{ \pi i k \Tr_{F} u^2} $ is the contribution of the $k$-level  CS term. The sum is over the
fundamental representation and it explicitly becomes Tr$_{F} u^2=\sum_i u_i^2$.
On the contrary the YM term does not contributes to the partition function;
\item 
the one loop contribution of the matter fields is  ${\det}_R e^{l(1-\Delta+i u)}$. The determinant is over the representation.
Explicitly it is
\be
 {\det}_R e^{l(1-\Delta+i u)} = \prod_{R}  e^{l(1-\Delta+i \rho_i(u))} 
\ee
where $\rho$ are the weights of $R$ in terms of the eigenvalues $u_i$.
The function 
$l(z)$
was introduced in \cite{Jafferis:2010un} (see also \cite{Hama:2010av} for an independent derivation).
It arises from the regularization of the one loop determinant of the matter fields. This one loop determinant is 
\be
\prod_{n=1}^{\infty} \left(\frac{n+1-\Delta+i \rho(u)}{n+1-\Delta-i \rho(u)} \right)^n
\ee
If there are not quantum corrections to the conformal dimension $\Delta$ the classical value $\Delta=1/2$ 
simplifies this expression and the regularized one loop determinant is  \cite{Kapustin:2009kz}
\be
\frac{1}{\cosh \rho_i(u)}
\ee
Anyway in this case we expect the R charge to couple with other flavor symmetry, and the classical result 
is quantum corrected. By using the Zeta functions the regularized function for the one loop determinant is
$e^{l(z)}$ where
\be
l(z) = -\frac{i \pi }{12}-z \text{Log}\left(1-e^{2 i \pi  z}\right)+\frac{1}{2} i \left(\pi  z^2+\frac{Li_2\left(e^{2 i \pi  z}\right)}{\pi }\right)
\ee
and $z=1-\Delta+i \rho_i(u)$.
\end{itemize}
In this paper we restrict to the $SU(N)_k$ cases in the $R+ \overline R$, with $R$ fundamental, adjoint, symmetric or  
antisymmetric.
In appendix \ref{sec6} we give some useful details for the weights of these representations.

The claim made in \cite{Jafferis:2010un} is that this partition function is extremized by the exact $R$ charge. 
This result is based on the assumption that  the partition function depends holomorphically  on
the combination $\Delta_j-i m_j$, where $m_j$ is a real mass term for the $j$-th chiral multiplet,
which conformal dimension is $\Delta_j$. From this holomorphy it follows that 
$\partial_\Delta  \mathcal{Z} \simeq \partial_m \mathcal{Z} $.
The one point function of an operator in a CFT over $S^3$ is  $\frac{1}{ \mathcal{Z}} \partial_m  \mathcal{Z}
 |_{m-0,\Delta=\Delta_{IR}}$,
and it vanishes at the conformal fixed point if parity is preserved. If parity is broken the 1-point function can be proportional to the identity.
The identity is a parity invariant operator, its VEV is real and Im$(1/ \mathcal{Z} \partial_{m_j} \mathcal{Z})$ vanishes. The result is that in this case the exact $R$ charge extremizes $| \mathcal{Z}|^2$. 

In \cite{Jafferis:2010un} this conjecture has been checked for cases with a one dimensional Cartan subalgebra. It was shown that 
R charge computed from the partition function agrees with the perturbative result of  \cite{Gaiotto:2007qi}. 

Here we check the conjecture 
for  $SU(N)_k$ CS matter theories at large $k$, and we observe that
the R charge computed from
the partition function agrees to the field theory expected result
even if the Cartan subalgebra has dimension higher than one.

\section{Field theory result} \label{sec2}

Here we briefly review the basic aspect of the field theory that we are considering,
discussed in  \cite{Gaiotto:2007qi}..
The theory is a $\mathcal{N}=2$ CS theory coupled with matter fields. 
The vector superfield V is composed by the gauge field $A_\mu$, the auxiliary scalars 
$\sigma$ and $D$ and the two component Dirac spinor $\chi$. 
The CS term in the action (in the WZ gauge) is
\be
S_{CS} = \frac{k}{4 \pi} \int \Tr \left(A \wedge d A +\frac{2}{3} A^3-\chi \overline \chi + 2 D \sigma \right) 
\ee
where $k$ is an integer denoting the CS level.
The matter field $\Phi=(\phi,\psi)$ is coupled to this vector multiplet by the action
\be
S_{\text{matter}} = \int D_\mu \overline \phi D^{\mu} \phi + i \overline\psi \gamma^{\mu} D_\mu \psi - \overline \phi \sigma^2 \phi
+\overline \phi D \phi - \overline \psi \sigma \psi + i \overline \phi \overline \chi \psi - i \overline \psi \chi \phi 
\ee
By integrating out $\chi $ and $D$ the authors observed that this action is not only classically but also quantum mechanically 
 marginal. 
 By adding a $\mathcal{N}=3$ coupling of $\Phi$ with $Q $ and $\tilde Q$, in the $R+\overline R$
 of the gauge group,
 they showed that the model reduces to a WZ model at large $k$.
 In the $\mathcal{N}=2$ case the coefficient of the two loops beta function is 
\begin{equation} \label{tracce}
b_0(R+\bar R) = \frac{2}{|R|} \left( \left(\Tr \, T^a T^b \right)^2 + \Tr \, T^a T^b T^a T^b \right)
\end{equation}
 and the $R$ charge for these matter fields is 
 \be
 R_{Q,\tilde Q} = \frac{1}{2} - \frac{b_0}{4 k^2} + \mathcal{O}\left(\frac{1}{k^4}\right)
 \ee
 where we are normalizing the trace of the fundamental such that $T(F)=1$.
The traces appearing in (\ref{tracce}) are
\bea
 \Tr \, T^a T^b   \Tr \, T^a T^b  &=& |G| \text{T}(r)^2  \non \\
 \Tr \, T^a T^b T^a T^b&=& |G| \text{T}(r) \left(\frac{\text{T}(r) |G|}{|r|} - \frac{\text{T}(G)}{2}\right)
\eea
In the case of $N_f$ flavor in the fundamental and antifundamental of $SU(N)_k$ 
we have 
\begin{equation}
R_{SU(N)_k}{(Q,\tilde Q)} = 1- \frac{(N_f N-1) \left(N^2-1\right)}{2 k^2 N^2}
\end{equation}
In the next sections we will show that the same result is obtained  by extremizing the partition function.

\section{Exact $R$ in $\mathcal{N}=2$ $SU(N)_k$ CS matter theories} \label{sec3}

In this section we explain general aspects of the computation of the exact $R$ charge from the extremization
of $\mathcal{Z}$.
We concentrate on representation $R+\overline R$, at large $k$, because we want to compare with the perturbative
field theory results.
By restricting to large values for the  CS level $k$ the matrix integral (\ref{partf}) is dominated by the saddle point $u_i=0$.
In the $SU(N)$ case there are $N-1$ independent variables $u_i$, because of the traceless condition.
The gaussian measure is  
\be
\text{Exp}\left(2 i \pi k \left( \sum_{i=1}^{N} u_i^2+ \sum_{i<j} u_i u_j \right)  \right)
\ee
It is possible to operate a coordinate transformation on the eigenvalues to diagonalize this
term,  but this procedure complicates the one loop contributions of the vector and the matter fields in terms of the 
eigenvalues.
We followed a different strategy: we integrate over $N$ and not $N-1$ eigenvalues, by imposing the traceless condition 
on the weights, Tr $u = 0$.
We show this procedure with an example. If we consider the gaussian integral
\be \label{pri}
\int_R e^{-2\alpha x^2} \text{d} x = \sqrt{\frac{\pi}{2 \alpha}} 
\ee
we can write this as
\be \label{sec}
\int_R \int_R e^{-2\alpha(x^2+y^2)} \delta(x+y) = \int_R \int_R e^{-2 \alpha(x^2+y^2)} \frac{1}{2 \pi}\sum_{m=-\infty}^{+\infty} e^{i m (x+y)}
\ee
where we used the Fourier expansion for the delta function
\footnote{Alternatively one can use the Fourier transform of the $\delta$ function and observe that the two results agree
because the integral receives contributions only at $u_i\simeq0$ .}.
The gaussian integrations on $x$ and $y$ can be separately performed by completing the squares
in the exponentials. 
By comparing the integral in (\ref{pri}) with the one in (\ref{sec}) we obtain
\be \label{use}
S_0 \equiv \sum_{m=-\infty}^{\infty} e^{-\frac{m^2}{ 2 \alpha}} = \sqrt{2 \pi \alpha}
\ee
We will use this procedure to compute the partition function for the $SU(N)$ gauge 
theory, by imposing the constraint with a delta function. Moreover we will
explicitly use the series $S_0$ and his derivative to respect to $\alpha$.
The partition function becomes 
\be
\mathcal{Z}\! = \!\sum_{m=\!-\infty}^{\infty} e^{\!-\!\frac{i N m^2}{4 k \pi}} \int \prod_{i=1}^N \text{d}u_i
e^{i \pi k \left(u_i+\frac{m}{2 k \pi}\right)^2}
\prod_{i<j}
\sinh^2(u_i\!-\!u_j) \,\,
{\det}_R e^{N_f (l(1+\Delta+i \rho(u))+ l(1+\Delta-i \rho(u)))}
\ee
Observe that the restriction to $R+\overline R$ simplifies the 
weight dependence of the partition function. Indeed here we 
use the relation $\overline \rho(u)=- \rho(u)$, i.e. the weights
of the conjugate representation $\overline R$ are opposite 
to the weights of $R$.

Then we shift the eigenvalues by a factor $\frac{m}{2 \pi k}$. This shift does not affect the
one loop contribution of the vector field. 
By expanding this term around $u_i \sim 0$ we have
\bea \label{kernel}
&&\Delta^2\left(\sinh(\pi u)\right) \simeq \Delta^2(\pi u) \prod_{i<j}\left(1+\frac{\pi^2}{3}(u_i-u_j)^2 +\frac{2\pi^4}{45}(u_i-u_j)^4 \right) 
 \\
 &&
\simeq \Delta^2(\pi u) \!\left(1+\frac{\pi^2}{3} \sum_{i<j} (u_i-u_j)^2 -\frac{\pi^4}{90} \sum_{i<j}(u_i-u_j)^4 + \frac{\pi^4}{18}
\left(
\sum_{i<j}(u_i\!-\!u_j)^2
\right)^2\
\right)\non
\eea
where we used the same strategy of \cite{Marino:2002fk,Aganagic:2002wv},  
computing the partition functions of matrix integrals with $\sinh^2 {\Delta}$ kernel.
Formula (\ref{kernel}) can be written in a more suitable way to perform the integration over the gauge group. The final result that we get is
\footnote{
Here we fix a convention which holds in all the paper. When a term
like $u_{i_A} u_{i_B}$ appears in the sum, if we don't' specify any condition (like $i_A<i_B$),
the constraint $i_A \neq i_B$ is understood.
For example the term $u_i u_j u_k u_l$ refers to the combination of four different weights.
}
\bea \label{vand}
&&\Delta^2\left(\sinh(\pi u)\right) \simeq\Delta^2\left( \pi u\right) \left(
1-\frac{\pi^2(N-1)}{3} \sum_{i=1}^{N} u_i^2-\frac{2 \pi^2}{3} \sum_{i<j} u_i u_j 
 \right.\non \\
 \!\!+\!\!&&\left.\frac{\pi^4(5 N^2\!-\!11 N\!+\!6)}{90} \sum_i u_i^4\!]!-\!\!
\frac{2 \pi^4(5N^2\!-\!6)}{45} \sum_{i<j}\left(u_i^3 u_j \!+\! u_i u_j^3 \right)
\!-\!\frac{2 \pi^2(N\!-\!3)}{9}
\sum_{j<k}u_i^2 u_j u_k 
\right.\non  \\
\!\!+\!\!&&\frac{\pi^4(5N^2-10 N+12)}{45}  \sum_{i<j} u_i^2 u_j^2
+\left.\frac{4\pi^4}{3}\sum u_i u_j u_l u_k
 \right)
\eea
All the terms in the parenthesis are  integrated over d$u_i$
by writing the Vandermonde determinant in terms of Hermite polynomials,
orthogonal to respect to the measure $e^{\pi i k u^2}$d$u$ \cite{metha}.
These polynomials are
\be
p_n(u) = \left(\frac{1}{i \pi k}\right)^n e^{-\frac{i \pi k u^2}{2}} \frac{\text{d}^n}{\text{d}u^n}
e^{\frac{i \pi k u^2}{2}}
\ee
normalized such that $h_n=\langle p_n,p_n \rangle$  is
\be
h_n = \frac{1}{\sqrt 2 \pi^{n+1}}  \left(-\frac{i}{k}\right)^{n+\frac{1}{2}} n!
\ee
The Vandermonde determinant in term of these polynomials is
$\Delta(\pi u)= \det\left( p_{N-1}(u)\right)$
or explicitly
\bea
\Delta(\pi u) &=& \det \left( 
\begin{array}{cccc}
1&1&\dots&1\\
\pi u_1 & \pi u_2 &\dots &\pi u_N \\
\dots& \dots&\dots&\dots \\
(\pi u_1)^{N-1} & (\pi u_2)^{N-1} &\dots  &(\pi u_N)^{N-1} 
\end{array}
\right) \non \\  &~&\\&=&  \det
\left( \begin{array}{cccc}
p_{0}(u_1) & p_{0}(u_2) & \dots & p_{0}(u_N) \\
p_{1}(u_1) & p_{1}(u_2) & \dots & p_{1}(u_N) \\
\dots &\dots &\dots &\dots \\
p_{N-1}(u_1) & p_{N-1}(u_2) & \dots & p_{N-1}(u_N) 
\end{array} 
\right) \non
\eea
The integration is performed by applying the recursion relation of the normalized
polynomials $P_n(u) = p_n(u)/\sqrt{h_n}$,
\be
u P_n(u) = \sqrt{r_{n+1}}P_{n+1}(u) + \sqrt{r_{n}} P_{n-1}(u)  
\ee
where $r_n=h_n/h_{n-1}=n/(i \pi k)$.
In the appendix \ref{sec7} we list all the integrals necessary to complete
the computation.
After the expansion of the Vandermonde we have to expand the 1-loop  determinant contribution
of the matter fields.  In the large $k$ limit we expand this term around $\Delta=1/2$ because 
we expect a small deviation from the classical result. From now on we define $\Delta=1/2+a$ and expand at $a\sim0$.

The shift on the eigenvalues modifies the function $l(z)$. Indeed in a generic representation 
a shift of a constant value $\xi$ on the eigenvalues modifies the weights from  $\rho(u)$ to $\rho(u+\xi)$
(an exception is the adjoint representation, as we already observed).
The effect of the shift becomes
\be
\rho(u) \rightarrow \rho(u+\xi) =  \rho(u) + T \xi 
\ee
because the weights are linear functions of $\rho(u)$;
$T$ represents the number of boxes in the Young tableaux associated to the representation
that we are considering, or equivalently the number of $u_i$ in the weights. In our case $\xi=\frac{m}{2 \pi k}$.
The one loop determinant is eventually expanded for large $k$ and for $u_i \sim 0$ and $a\sim0$.
This expansion is formally written as 
\be \label{gen}
\prod_{i=1}^N e^{ N_f\left(l\left(1-\Delta+i \left(u_i-\frac{m}{2 k \pi}\right)\right)+l\left(1-\Delta - i \left(u_i-\frac{m}{2 k \pi}\right)\right)\right)}=
\sum_{i,j} \sum_{\alpha,\beta}
f_{\alpha \beta \gamma}(a)  u_i^\alpha u_j^\beta u_{(m)}^{\gamma}
\ee
The $u_i$ dependence is determined by the knowledge of 
 $\sum \rho_i^2$ and $\sum \rho_i^4$,
 where the sums run only over the weights of $R$,
if the representation is $R+\overline R$ .
Moreover we defined $u_{(m)}=m/k$, because this term has a similar role than $u_i$ in the large $k$ expansion.
Indeed the $u_m$ expansion adds some powers on $m^2$, which modify some of the series over $m$. The modified
series are
\be
S_2 = \sum_{m=-\infty}^{\infty} \frac{m^2}{k^2}  e^{-\frac{i N m^2}{4 k \pi}}=-\frac{ 4 \pi i}{N} \partial_{k} S_0
\, , \,
S_4 = \sum_{m=-\infty}^{\infty} \frac{m^4}{k^4}  e^{-\frac{i N m^2}{4 k \pi}}= -\frac{16 \pi^2}{N}\left(\partial_k^2+\frac{2}{k}\partial_k\right)S_0
\ee
By computing the relevant terms in the product of  (\ref{vand}) and (\ref{gen})
and by using the integrals in the appendix \ref{sec7} and the series $S_0,S_2$ and $S_4$ 
we obtain the partition function.
The last step for finding the exact $R$ charge is then solving the equation $\partial_a |Z|^2=0$.
In the rest of this section  we apply this procedure top many representation of $R+\overline R$,
showing the agreement of this technique and the expected perturbative results.

\subsection{Fundamental and antifundamental representations}

Here we compute the partition function for a generic $SU(N)_k$ gauge theory
 with $N \ll k$ and with $N_f$ fundamental and $N_f$ antifundamental quarks
 \footnote{In \cite{Jafferis:2010un} the author used a different convention, $N_f^{his}=N N_f^{our}$. }.
The partition function is
\be
\mathcal{Z}  = \int \prod_{i=1}^{N} \text{d}u_i e^{i \pi k u_i^2+ N_f(l(1-\Delta+i u_i)+l(1-\Delta - i u_i))}\prod_{i<j}\sinh^2(u_i-u_j)
\delta(u_1+\dots+u_n)
\ee
As we anticipated above we write the delta function in terms of his Fourier series and then we make a change of variables,
by completing the squares in the exponential. The partition function that we obtain is
\be
\mathcal{Z} \! = \!\sum_{m=\!-\infty}^{\infty} e^{\!-\!\frac{i N m^2}{4 k^2 \pi}} \int \prod_{i=1}^N \text{d}u_ie^{i \pi k u_i^2\!+\! N_f\left(l\left(1\!-\!\Delta+i \left(u_i\!-\!\frac{m}{2 k \pi}\right)\right)+l\left(1\!-\!\Delta \!-\! i \left(u_i\!-\!\frac{m}{2 k \pi}\right)\right)\right)}\prod_{i<j}\sinh^2(u_i\!-\!u_j)
\ee
We then  factor out the Vandermonde determinant $\Delta^2(\pi u)$ as in (\ref{vand}), 
and we write it in terms of orthogonal Hermite polynomials.
The 1 loop determinants are expanded as in (\ref{gen}). 
Explicitly the weights are $\rho_i=u_i$ and 
the $u_i$ dependence of the expansion is 
\be \sum_{Fund.} \rho^2 = \sum_{i=1}^{N} u_i^2 \quad\quad,\quad\quad \sum_{Fund.} \rho^4 = \sum_{i=1}^{N} u_i^4 \ee
The partition function at the leading order in $a$ at large $k$ 
becomes
\footnotesize
\bea
\mathcal{Z} &\simeq&
288 k^2 N (2+a^2 N_f N \pi ^2)+24 i \pi  k N (N^2-1) (4 N+N_f(a (24+a (N^2-3 N_f N\!-\!6) \pi ^2)\!-\!6))\non \\&-&
(N^2\!-\!1) \pi ^2 (8 N^5\!+\!18 (4 a\!-\!1)^2 N_f^2 N (N^2\!+\!1)\!-\!12 N_f(3\!+\!2 N^4 \!- \!8 a (N^4\!-\!N^2\!+\!3))\!-\!a^2 N_f(144\non 
\\&-&N (9 N_f^2 N(N^2+1)+6 N_f(2 N^4\!-\!6N^2\!-\!3)\!-\!4 N (N^4-6 N^2+18))) \pi ^2) \non\\
\eea
\normalsize
where we omitted the overall factors, which do not modify the 
extremization.
By extremizing $|\mathcal{Z} |^2$ to respect to $a$ we finally obtain
\be
R=\
\Delta=\frac{1}{2}-a=\frac{1}{2}-\frac{(N_f N-1)(N^2-1)}{2 k^2 N^2}
\ee
exactly as expected from the perturbative computation.

\subsection{Adjoint representation}

Here we discuss the case of matter fields in the adjoint representation. By observing that the adjoint is self conjugate 
$R_{Adj} = R_{\overline {Adj}}$  the $g$ adjoints fields can be thought as $g/2$ fields and $g/2$ conjugate fields
\footnote{We restrict to an even number of adjoint, such that we can use the formula (\ref{tracce}) as explained in
\cite{Gaiotto:2007qi}.}.
By applying this observation to the field theory computation we expect that for $g$ adjoint the $R$ charge is 
\be \label{Adjexp}
R = 1 - \frac{N^2(g+1)}{k^2}~~.
\ee
In this case the shift introduced by the delta function does not affect the integrations
because the weights appearing in the one loop determinant are in the adjoint representation.
Indeed the shift  
\be
u_i \rightarrow u_i - \frac{m}{2 \pi k}
\ee
gives an overall factor, but cancels in the differences $u_i-u_j$.

The rk$_G$ zero-weights of the adjoint contribute to the partition function.
In  $SU(N)_k$  they are $N-1$ and they add to the 1-loop determinant 
a factor rk$_G \times l(1-\Delta)$. 
The partition function for the $SU(N)_k$
gauge theory  with $g$ adjoint matter fields is
\be
\mathcal{Z} \simeq 
\int \prod_{i=1}^{N} \text{d}u_i ~e^{i \pi k u_i^2}
\prod_{i<j} \sinh^2\left(\pi(u_{ij})\right)
\left( e^{ g \left(l(1-\Delta + i u_{ij})+ l(1-\Delta -i u_{ij} )\right)} \right)
 e^{g(N-1)l(1-\Delta)}
\ee
where $u_{ij}=u_i-u_j$, and we omitted the overall series introduced
by the delta function.
The one loop determinant for the vector field is expanded as above.
Moreover the weights $\rho(u)=u_i-u_j$ contribute to the one loop determinant of the 
matter field as
\be
\sum_{Adj.} \rho^2(u)= (N-1) \sum_{i=1}^{N} u_i^2 -2 \sum_{i<j} u_i u_j 
\ee
and
\be
\sum_{Adj.} \rho^4(u) = (N-1) \sum_{i=1}^{N}  u_i^4+6 \sum_{i<j} u_i^2 u_j^2 + 4 \sum_{i<j} (u_i^3 u_j+u_i u_j^3)
\ee
The partition function is proportional to
\footnotesize
\bea
&&
288 k^2 (4+a^2 g (N^2-1) \pi ^2)+24 i \pi k N (N^2-1)  (8+g (a (48+a (3 g(1-N^2)+2 N^2-14) \pi ^2)-12 )) \non \\ \!&\!-\!&\!
\pi ^2N^2 (N^2\!-\!1) (12 g (6\!-\!64 a+3 (4 a\!-\!1)^2 g)+4 (2+3 (4 a-1) g)^2 N^2+a^2 g (9 (48-(g-6) g)+2 (51 g \non \\
\!&\!-\!&\! 26) N^2+(3 g-2)^2 N^4) \pi ^2) \non\\ 
\eea
\normalsize
where we omitted the overall factor.
The $R$-charge obtained from the extremization of this partition function coincides with 
(\ref{Adjexp}).

\subsection{Symmetric representation}

Here we consider a matter field in the $M$-index symmetric  representation 
and its conjugate. 
The weights are 
\be
\rho_i(u) = u_{i_1}+\dots + u_{i_M}
\ee
without any restrictions on the $u_i$.
We compute the sums $\sum \rho^2$ and  $\sum \rho^4$ of the weights  in terms of the eigenvalues
$u_i$. By defining $C_{i}^{j}=\left(\begin{array}{c} i\\ j \end{array}\right)$$\begin{array}{c}~\\~\\~\\\end{array}\!\!\!$ we obtained
\bea
\sum_{Sym.} \rho^2 = \sum_{i=0}^{M} (M-i)^2
 C_{N+i-2}^{i} \sum_{i=1}^{N}  u_i^2 
 +
 2 \sum_{i=1}^{M-1} (M-i)
 C_{N+i-2}^{i-1}
 \sum_{i<j} u_i u_j 
 \eea
where the coefficients in the sum over $i$ can be written in terms of the dimension of the
representation $|R|$ as
\bea  \sum_{i=0}^{M} (M-i)^2
 C_{N+i-2}^{i}  = 
 \frac{M(N+2M-1)}{N(N+1)} |R|
\,\,
 ,
\, \,
 \sum_{i=1}^{M-1} (M-i)
 C_{N+i-2}^{i-1}=
 \frac{M(M-1)}{N(N+1)} 
 |R| \non \\
\eea
The expansion of $\sum \rho^4$ is
\bea
&&\sum_{Sym.} \rho^4 =
\sum_{i=0}^{M} (M-i)^4
C_{N+i-2}^{i}  \sum_{i=1}^{N}  u_i^4
 + 4
\sum_{i=1}^{M-1} (M-i)^3
C_{N+i-2}^{i-1} 
\sum_{i<j}\left(u_i^3 u_j + u_i u_j^3\right) \non \\
&&+
6  \sum_{i=1}^{M-1}
  (\!M\!-\!i)^2
  ( C_{N+i-2}^{i-1}\!\!+\!2C_{N+i-2}^{i-2})
  \sum_{i<j}u_i^2u_j^2 \!+\!
  12 \sum_{i=2}^{M-2} 
  (\!M\!-\!i)^2 C_{N+i-2}^{i-2}
  \sum_{j<k} u_i^2 u_j u_k \non \\
&&+
  24 \sum_{i=3}^{M-3} (M-i) C_{N+i-2}^{i-3}
  \sum u_i u_j u_l u_k
\eea
Even in this case  the coefficients can be written in terms of $|R|$.
The partition function is proportional to
\footnotesize
\bea
&&
288 k^2 N^2 (a^2 |R| \pi ^2\!+\!2)(N+1)+24 i k \pi N (N^2-1)  (4 N^2 (N+1)+6 (4 a\!-\!1) |R| M (M+N)
-
a^2 |R| \non \\
&&  (3 ( |R|\!+\!2) (M\!+\!N)M \!-\!2 N^2 (N\!+\!1)) \pi ^2)\!-\!(N\!-\!1) \pi ^2 (8 N^6 (N\!+\!1)^2\!+\!9 a^2 
 |\!R|^3 M^2 (M\!+\!N)^2 (N^2\!+\!1) \pi ^2 \non \\ &&
 - 4 |\!R| (N\!+\!1) (3 M (M\!+\!N) (3 (8 a\!-\!1\!) M (N\!\!-\!1) (M\!+\!N)
\!+\! N^2 (3\!+\!2 N^2\!\!-\!8 a (N^2\!\!+\!2)))\!+\!a^2 (6 M^2 (N\!\!-\!\!3)^2 \non \\&&
N^2\!\!-\!\!N^6 (N\!+\!1)\!-\!36 M^3 (N\!\!-\!\!1) 
(M\!+\!2 N) 6 M N^3 (N^2\!+\!3)) \pi ^2)\!+\!6 |R|^2 M (M\!+\!N) (3 (4 a\!\!-\!\!1)^2 M (M\!+\!N)\non \\
&& (N^2\!+\!1)
+a^2 (3M (M+N)(3 N^2+1)-N^2 (N+1) (2 N^2+3)) \pi ^2)) \non \\
\eea
\normalsize
where we are omitting the overall term.
By extremizing $|\mathcal{Z} |^2$ to respect to $a$ we found
\be
R=\frac{1}{2}-a=\frac{1}{2}- \frac{M(N-1)(M+N)((N+1)M^2 + M (N+1)(|R|+|G|))}{2 k^2 N^2(N+1)}
\ee
By observing that for a $M$-index symmetric representation
\be
T(r) =  \frac{M(N+M)}{N(N+1)} |R|
\ee
our result agrees with the two loop field theory computation.

\subsection{Antisymmetric representation}

Here we compute the partition function for 
matter fields in the $L$ index antisymmetric representation and 
its conjugate $\overline L = N-L$.
The weights $\rho(u)$ are 
\be
\rho(u) = u_{i_1} + \dots + u_{i_L} \quad \quad \text{with}\quad  i_j\neq i_k \quad \text{if} \quad  j \neq k
\ee
The $u_i$ dependence of the one loop determinant for the antisymmetric matter field is 
is computed from the relation
\be
\sum_{Antisym.} \rho^2(u) = C_{N-1}^{L-1} \sum u_i^2 + 2 C_{N-2}^{L-2} \sum_{i<j} u_i u_j
\ee
and
\bea
\sum_{Antisym.} \rho^4(u) &=& 
 C_{N-1}^{L-1} \sum u_i^4+
4 \, C_{N-2}^{L-2}  \sum_{i<j} \left( u_i^3 u_j+u_i u_j^3\right) +
6 \, C_{N-2}^{L-2}     \sum_{i<j} u_i^2 u_j^2\non \\&+&
  12 \, C_{N-3}^{L-3}  \sum_{j<k} u_i^2 u_j u_k+
   24 \, C_{N-4}^{L-4}  \sum u_i u_j u_k u_l
\eea
Even in this case we can write the binomial coefficients in terms of the 
dimension of the representation. We then computed the
partition function, it is proportional to
\footnotesize
\bea
&&
288 k^2 N^2 (a^2 |R| \pi ^2+2)(N-1)+24 i k \pi  N(N^2-1) (4 (N-1) N^2+6 (4 a-1) |R| 
L (N\!-\!L)-a^2 |R| \non \\&& 
(3 (|R|\!+\!2) L(N\!-\!L)\! -\!2 (N\!-\!1) N^2) \pi ^2)\!-\!(N\!+\!1) \pi ^2 (8 (N\!-\!1)^2 N^6\!+9 a^2 |R|^3 \
L^2 (N-L)^2 (N^2\!+1) \pi ^2 \non \\&&
-4 |R| (N-1) (3 L (N\!-\!L) (3 (8 a\!-\!1) L(N\!-\!L) (N\!+\!1)\!+\!
N^2 (2 N^2\!+\!3\!-\!8 a (N^2+2)))\!+\!a^2 ((N-1) N^6 \non \\&&-
6 L^2 N^2 
(N\!+\!3)^2\!-\!6 L N^3 (N^2\!+\!3)) \pi ^2) \!+\!6 |R|^2 L (N\!-\!L) (3 (4 a\!-\!1)^2 L (N\!-\!L) (N^2\!+\!1)-
a^2 ((N-1) \non \\&&N^2 (2N^2+3)+3L(N-L) (3 N^2+1)) \pi ^2))\non \\
\eea
\normalsize
By extremizing $|\mathcal{Z} |^2$ to respect to $a$ we found
\be
R=\frac{1}{2}-a=\frac{1}{2}- \frac{L(N+1)(N-L)((N-1)L^2 - L (N-L)(|R|+|G|))}{2 k^2 N^2(N-1)}
\ee
which agrees with the field theory computation once we observe that
\be
T(r) = \frac{L(N-L)}{N(N-1)}|R|
\ee

\section{Conclusions}

In this paper we checked that the extremization of the partition function of $\mathcal{N}=2$ 
three dimensional YM CS theories leads to the correct $R$ charge, by matching the results of $SU(N)_k$ theories with
the perturbative  results.
We explicitly computed the partition function
and the $R$ charge for the fundamental, adjoint, symmetric and antisymmetric representations.
 It would be interesting to check this agreement for other representations and gauge groups.

Another interesting problem is the application of this procedure to $\mathcal{N}=2$
CS matter theories conjectured to be dual to M theory on AdS$_4 \times $SE$_7$, where SE$_7$ 
is a seven dimensional Sasaki
Einstein manifold \cite{conj}. Indeed in that case the exact $R$ symmetry 
is known from the minimization of the volumes. In \cite{Martelli:2011qj} this study has been 
started, and the authors showed  that at large $N$ the results agree for some two and three gauge group theory. 
It would be interesting extend the computation to theories with more gauge groups and check 
if the partition functions coincide for the proposed Seiberg and toric dualities of \cite{duality}.

A last observation concerns the existence of a $c$-theorem in three dimensions. 
In four dimensions the Cardy conjecture \cite{Cardy:1988cwa} argues that the central charge is the
analogous of the Virasoro central charge 
of two dimensional gauge theories \cite{Zamolodchikov:1986gt}. 
It takes into account the reduction of the massless degrees of freedom of the theory between
the UV and the IR fixed point.
In its strongest version the conjectured $a$-theorem 
states that $a$ is a positive monotonic decreasing function in an RG flow from the UV to the IR.
Many checks of this conjecture exist,
confirming the prediction of a monotonic decreasing behavior of $a$
during the RG flow, by
using some refined version of $a$-maximization out of the fixed point
\cite{Kutasov:2003ux}.
Here we observe that in all the examples we computed the 
extremized value of the $R$ charge is a minimum of the partition function. 
By supposing that the actual procedure of finding the exact $R$ charge 
is the minimization of the partition function, this last quantity may play the role of the
central charge $a$ (or its inverse) in four dimension, and a 
 $c$-theorem may be formulated (see \cite{Jafferis2}
for a recent discussion on this possibility).

\section*{Acknowledgments}
It is a great pleasure to thank Ken Intriligator
for many helpful discussions and clarifications. A.A.
is supported by UCSD grant DOE-FG03-97ER40546.

\appendix

\section{Conventions on Lie Algebra}  \label{sec6}

In this appendix we review the conventions  on Lie algebra that we used to perform the computation.
We computed the partition functions in terms of the weights of the fundamental representation. 
The weights of the other representations are expressed in terms of these weights.
We choose the following basis for the weights of the fundamental representation
\begin{equation} \label{wf}
u_1=(1,0,\dots,0), \quad u_2 =(-1,1,0,\dots,0),\quad \dots\quad u_{N}=(0,\dots,-1)
\end{equation}
with $N-1$ elements in each weight.
The weights of the antifundamental are the same with opposite sign.
The highest weight, $u_1$ is associated to the $A_{N-1}$ Dynkin diagram
describing $SU(N)$. 
There is a correspondence between the highest weight  and the representations, because
the $i-th$ element $N_i$ in the highest weight is the number of column of  length
$N_i$ in the Young tableaux.
A generic representation is always described by its highest weight, 
and its knowledge is enough to construct all the other weights.
These other weights are computed by acting with the raws of the Cartan matrix 
on the highest weight.
\begin{figure}
\begin{center}
\includegraphics[width=12cm]{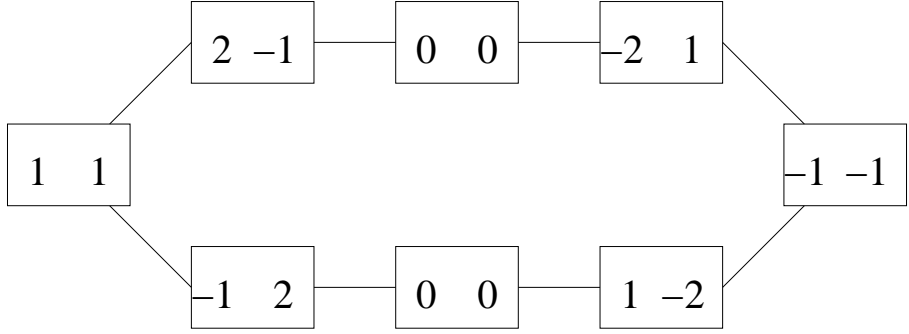}
\caption{Weights of the Adjoint of $SU(3)$}
\label{adjsu3}
\end{center}
\end{figure}
Moreover it is possible to express the raws of the Cartan matrix to respect of the weights of the 
fundamental representation. Indeed they are simply $u_{i}-u_{i+1}$.
By subtracting these quantities to the highest weight of a generic representation 
the other weights are computed.
If we consider the adjoint representation its highest weight is 
$(1,0,\dots,0,1)$, which in terms of the fundamental representation is $u_1-u_n$,
the action of $u_i-u_{i+1}$ gives the usual $u_i-u_j$ $N(N-1)/2$ non zero weights
and in addiction there are $N-1$ zero weights.
For example for $SU(3)$ the highest weight is $u_1-u_3=(1,1)$. Then we subtract 
the raws $u_1-u_2=(2,-1)$ and $u_2-u_3=(-1,2)$ obtaining the other 
weights. Usually it is not necessary to introduce this specific basis for
the $u_i$ and the weights of the representation are derived as in 
figure \ref{adjsu3}. Anyway here we choose the basis $u_i$ because it is
more convenient for the integrations over the gauge group.
In the paper we studied also the M-index symmetric representation $(M,0,\dots,0)$
(with his conjugate $(0,\dots,M)$)
and the L-index antisymmetric $(0,\dots,1_L,\dots,0)$ ( with its conjugate).
We computed the weights for these representations by starting from the highest one, 
and finally we expressed the sum 
$\sum \rho^2$ and $\sum \rho^4$ in term of (\ref{wf}). 

\section{Integrals} \label{sec7}
In this appendix we list the integrals that we used to perform the computation
for the $U(N)_k$ theory.
The generic integral is 
\be
I = \int 
\prod_{i}\text{d}u_i e^{i k \pi \Tr u^2}f(\{u_i\}) \Delta^2(\pi u) 
\ee
If $f(\{u_i\})=1$ we obtain 
\be
 I_1 = \sqrt{(-1)^N} e^{-\frac{1}{4} i N^2 \pi } k^{-\frac{N^2}{2}} 
\left(\frac{\pi }{2}\right)^{\frac{1}{2} (N-1) N} \text{G}_2 (N+2) 
\ee
where G$_2$ is the Barnes function defined by G$_2(z+1)=\Gamma(z) $G$_2(z)$,
and G$_2(1)=1$.
This integral differs from the usual integral of the Vandermonde by a power of $2^{N(N-1)}$,
because we omitted a factor of $2$ in front of $\sinh^2(\pi (u_i-u_j))$
in the original partition function. 
For different values of the function $f(\{u_i\})$ the integrals are computed 
from the recursive relation of the orthogonal polynomials.
The integrals that are necessary 
for our computation are listed below.
\be
\begin{array}{|l||c|}
\hline
f(\{u_i\})&I/I_1\\
\hline
\sum u_i^2
&
\frac{i N^2}{2 \pi k}
\\
\sum_{i<j} u_i u_j
&
-\frac{i N(N-1)}{4 \pi k}
\\
\sum_{i} u_i^4&
-\frac{N(2 N^2+1)}{4 \pi^2 k^2}
\\
\sum_{i<j} u_i^2 u_j^2
&
-\frac{N(N-1)(N^2-N+1)}{8 \pi^2 k^2}
\\
\sum_{j<k} u_i^2 u_j u_k
&
\frac{N(N-1)(N-2)^2}{8 \pi^2 k^2}
\\
\sum_{i<j}(u_i^3 u_j + u_j^3 u_i)
&
\frac{N(N-1)(2 N-1)}{4 \pi^2 k^2}
\\
\sum u_i u_j u_k u_l 
&
-\frac{N(N-1)(N-2)(N-3)}{32 \pi^2 k^2}\\
\hline
\end{array}
\ee

\end{document}